\documentclass[twocolumn,prl,aps,nobalancelastpage,amsfonts]{revtex4}

\usepackage{graphicx}
\begin{document}

\title{Temperature dependent anisotropy of the penetration depth \\and coherence length in MgB$_2$}
\author{J.D. Fletcher, A. Carrington, and O.J. Taylor}
\affiliation{H.H. Wills Physics Laboratory, University of Bristol, Tyndall Avenue, BS8 1TL, United Kingdom.}
\author{S.M. Kazakov and J. Karpinski}
\affiliation{Laboratorium f\"{u}r Festk\"{o}rperphysik, ETH Z\"{u}rich, CH-8093 Z\"{u}rich, Switzerland.}
\date{\today}

\begin{abstract}
We report measurements of the temperature dependent anisotropies ($\gamma_\lambda$ and $\gamma_\xi$) of both the London
penetration depth $\lambda$ and the upper critical field of MgB$_2$. Data for $\gamma_\lambda=\lambda_c/\lambda_a$ was
obtained from measurements of $\lambda_{a}$ and $\lambda_c$ on a single crystal sample using a tunnel diode oscillator
technique. $\gamma_\xi=H_{c2}^{\parallel c}/H_{c2}^{\bot c}$ was deduced from field dependent specific heat
measurements on the same sample. $\gamma_\lambda$ and $\gamma_\xi$ have opposite temperature dependencies, but close to
$T_c$ tend to a common value ($\gamma_\lambda\simeq \gamma_\xi=1.75\pm0.05$).  These results are in good agreement with
theories accounting for the two gap nature of MgB$_2$.
\end{abstract}
\maketitle

The existence of two superconducting gaps in MgB$_2$ \cite{MazinA03} leads to an unusually strong temperature
dependence in the anisotropy of the upper critical field, $\gamma_\xi=H_{c2}^{\parallel c}/H_{c2}^{\bot c}$ (Refs.\
\onlinecite{SologubenkoJKKO02,AngstPWJKKRK02,Kogan02,DahmS03}). The maximum value of $\gamma_\xi\simeq6$ is found at
the lowest temperatures where it is dominated by the anisotropy in the quasi two dimensional $\sigma$ bands. At higher
temperatures, $\gamma_\xi$ is substantially reduced due to the increasing contribution of the more isotropic $\pi$ band
\cite{DahmS03}. The anisotropy of the penetration depth, $\gamma_\lambda=\lambda_c/\lambda_{a}$, is predicted to have a
markedly different temperature dependence to that of $H_{c2}$ \cite{Kogan02,GolubovBDKJ02}. At zero temperature, in the
clean local limit, the anisotropy in the penetration depth is determined only by the anisotropy in the Fermi velocity
$v_F$ and not by the anisotropy in the gap \cite{ChandrasekharE93}. Band structure calculations \cite{KortusMBAB01}
show that the average $v_F$ is approximately isotropic suggesting that $\lambda$ is similarly isotropic at low
temperature \cite{Kogan02}. As temperature is increased it has been predicted \cite{Kogan02,GolubovBDKJ02} that
$\gamma_\lambda$ increases markedly because of the more rapid reduction of the superfluid density on the $\pi$ sheets
(where the gap is smaller) than on the $\sigma$ sheets.

There have been several experimental measurements of the temperature dependence of $\gamma_\xi$. Although there were
some disagreements in early measurements, the most recent data
\cite{SologubenkoJKKO02,AngstPWJKKRK02,RydhWKKCBLKMKKKLL04} agree well with the theoretical predictions
\cite{Kogan02,DahmS03}. Significantly less experimental data has been reported on $\gamma_\lambda$. Measurements of the
distortion of the vortex lattice by neutron scattering \cite{CubittEDJKK03} and scanning tunneling spectroscopy
\cite{EskildsenJLKFJKK03} suggest that at low temperature the anisotropy is small, $\gamma_\lambda=1.2\pm 0.1$.
$\gamma_\lambda$ has also been estimated from the anisotropy of $H_{c1}$ ($\gamma_{H_{c1}}=H^{\|c}_{c1}/H^{\bot
c}_{c1}$). There has been some disagreement between different studies (see Ref.\ \onlinecite{LyardSKMMKKLL04}) but
recent data \cite{LyardSKMMKKLL04} suggest that at low $T$, $\gamma_{H_{c1}}\simeq 1$ and there is an upward trend in
$\gamma_{H_{c1}}$ near $T_c$.  However, $\gamma_{H_{c1}}$ may not be simply related to $\gamma_\lambda$ in MgB$_2$
because of two gap effects.  In this paper, we present direct measurements of the temperature dependencies of
$\lambda_{a}$ and $\lambda_{c}$ in single crystal samples of MgB$_2$ in the Meissner state, using a sensitive radio
frequency technique. Our data show that $\gamma_\lambda$ increases as $T\rightarrow T_c$, in agreement with theoretical
predictions. Measurements of $H_{c2}$ anisotropy in the same crystals show that the values of $\gamma_\lambda$ and
$\gamma_\xi$ converge as $T\rightarrow T_c$.

Single crystals of MgB$_2$ were grown using a high pressure technique described elsewhere
\cite{KarpinskiAJKPWRKPDEBVM03}. Some crystals used in this study are from the same batch as those used in de Haas-van
Alphen studies \cite{Fletcher}, and are hence known to be of high quality, with mean free paths of $\ell_\pi \simeq
850$~\AA, $\ell_\sigma \simeq 500$~\AA. The $T_c$ of the crystals, as determined from the midpoint of the heat capacity
anomaly, was 38.3~K. Measurements of the temperature dependent penetration depth were performed using a tunnel diode
oscillator technique operating at 11.7~MHz \cite{CarringtonGKG99}. Samples  were mounted on a temperature controlled
sapphire stage (base temperature 1.5~K) with the probe field aligned either into, or along the boron planes.

Changes in the circuit resonant frequency $\Delta F$ are directly proportional to the change in penetration depth
$\Delta \lambda$ as the sample temperature is varied. With $H\parallel c$ the shielding currents flow only in the basal
plane, and hence the measured frequency shifts are directly proportional to $\Delta \lambda_{a}$. The constant of
proportionality is determined from the geometry of the sample \cite{ProzorovGCA00} with an accuracy of  $\sim5-10$\%.
For $H\bot c$ the shielding currents flow both along the $c$-axis and in the basal plane and the measured frequency
shifts contain contributions from both $\Delta \lambda_a$ and $\Delta \lambda_c$. In a rectangular sample
\cite{ProzorovGCA00} \begin{equation} \frac{\Delta F^{\bot c}}{\Delta F_0^{\bot c}} = \frac{\Delta\lambda_{a}}{t} +
\frac{\Delta\lambda_c}{w}, \label{eqLambdaEff}\end{equation} where $2t$ and $2w$ are the sample dimensions in the $c$
direction and in the in-plane direction (perpendicular to the field) respectively. $\Delta F_0^{\bot c}$ is the
frequency shift observed when the sample is withdrawn from the coil, and accounts for the demagnetizing factor of the
sample. For this study samples with simple rectangular shapes and small aspect ratios were selected so as to maximize
the $c$-axis contribution.  The above formula neglects any contribution from the top and bottom faces of the sample.
For crystals with aspect ratios equal to those reported here we expect this to be a small correction ($\lesssim 5\%$).
This is supported by the fact that experimentally we find very similar values of $\gamma_\lambda$ for samples with
different aspect ratios. Most of the data presented here are for a crystal with dimensions
(0.44$\times$0.33$\times$0.15)mm$^3$ with the shortest direction being along the $c$-axis. The accuracy of alignment
with the field is estimated to be better than $5^\circ$ in all cases. Corrections due to misalignment scale with
$\sin^2\theta$, and are small \cite{ProzorovGCA00}. The estimated uncertainty in the absolute values of $\Delta \lambda
_{a}$ is 10\%, and 20\% for $\Delta \lambda_c$.

\begin{figure}
\center
\includegraphics*[width=7cm]{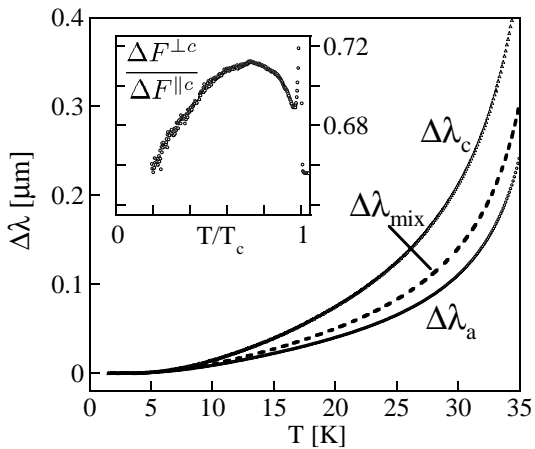}
\caption{Temperature dependence of $\Delta\lambda_a$, $\Delta \lambda_{c}$ and $\Delta \lambda_{\rm mix}$ for a single
crystal of MgB$_2$. Inset: Ratio of the measured frequency shifts for $H\|c$ and $H\bot c$.} \label{figlam}
\end{figure}

In Fig. \ref{figlam} we show the measured temperature dependencies of both $\Delta\lambda_{a}$ and $\Delta\lambda_c$.
In this figure we also show $\Delta \lambda_{\rm mix}=t\Delta F^{\bot c}/\Delta F_0^{\bot c}$. This quantity is
directly proportional to the raw frequency shift and would equal $\Delta \lambda_a$ if the contributions from the
$c$-axis currents were negligible. The presence of the $\Delta\lambda_c$ contributions enhances $\Delta \lambda_{\rm
mix}$ with respect to $\Delta\lambda_{a}$. We find that the temperature dependence of $\Delta\lambda_c$ is similar to
$\Delta\lambda_a$, but a factor $\sim$ 2 larger. This is consistent with the behavior reported previously
\cite{ManzanoCHLYT02}, although in that study the crystals used were not of sufficiently uniform thickness to
accurately determine $\lambda_c$.

\begin{figure}
\center
\includegraphics*[width=7cm]{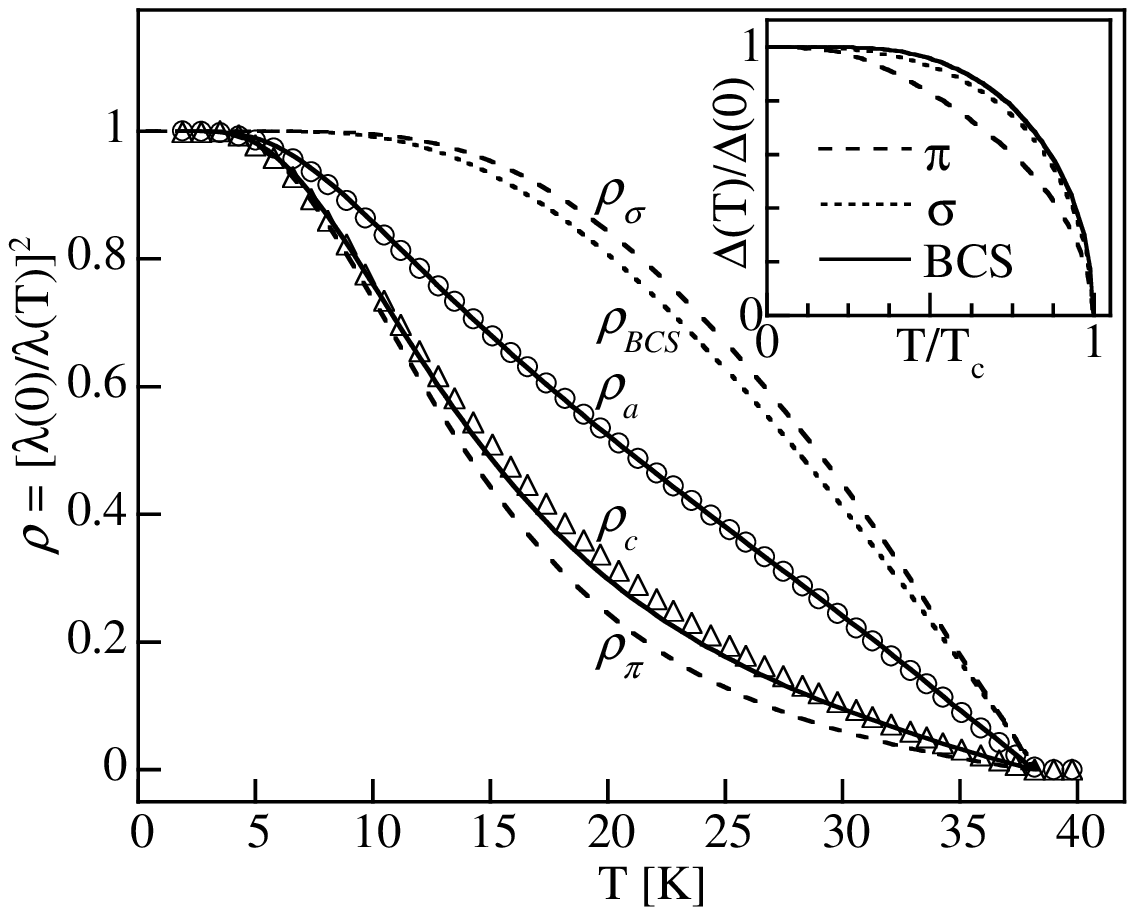}
\caption{In-plane and interplane superfluid density, $\rho_a$ and $\rho_c$ calculated from $\Delta \lambda (T)$, along
with fits to the two gap model (solid lines).  The behavior of $\rho_\sigma$ and $\rho_\pi$ deduced from the fits is
also shown (dashed lines). The dotted line shows the isotropic BCS weak coupling behavior. Inset: Temperature dependent
gap functions used in the fit: (dashed and dotted lines) taken from Ref.\ \onlinecite{BrinkmanGRDKKJA02}. The BCS gap
function (solid line) is also shown.} \label{figrho}
\end{figure}

The penetration depth data are more easily interpreted by calculating the normalized superfluid density
$\rho=[\lambda(0)/\lambda(T)]^2$.  For these calculations we take $\lambda_a(0)=\lambda_c(0)= 1000$~\AA~in accord with
$\mu$SR and neutron studies \cite{NiedermayerBHKA02,CubittLBAC03,OhishiMAKHK03}.  We will discuss the effect of varying
$\lambda_a(0)$ and $\lambda_c(0)$ later.

The calculated superfluid densities $\rho_{a}(T)$ and $\rho_c(T)$ are shown in Fig. \ref{figrho}. $\rho_a(T)$ and
$\rho_c(T)$ have quite different temperature dependencies and are both considerably different to that expected for a
conventional isotropic BCS superconductor. At low temperature both superfluid densities approach unity exponentially,
but at higher temperature they have a much stronger temperature dependence than that expected in the conventional case.

Previously \cite{ManzanoCHLYT02}, it was shown that $\rho_a(T)$ can be well described by a phenomenological two gap
model, which was first applied to MgB$_2$ by Bouquet \textit{et al.} \cite{BouquetWFHJJP01}.  In this model, the
contributions to the superfluid density from the $\sigma$ and $\pi$ bands ($\rho_\sigma$ and $\rho_\pi$) are calculated
independently using a temperature dependent energy gap $\Delta_{k}(T)$ which follows the usual weak coupling BCS form
but has a modified zero temperature value. These two superfluid densities are then added to give the total
\begin{equation} \rho_{i}(T) = x_{i}\rho_{\pi}(T) + (1-x_{i})\rho_{\sigma}(T)
\end{equation} here $i$ refers to the crystal direction ($a$ or $c$). The parameter $x_i$ sets
the relative contribution of the $\pi$ band to the total normalized superfluid density.  Theoretically $x_i$ is related
to band structure by \cite{ChandrasekharE93}
\begin{equation}
x_i=\frac{X_i^\pi}{X_i^\sigma+X_i^\pi},\quad X^{k}_i=\int \frac{(v_i^k)^2}{v_F} dS_k \label{xeq}
\end{equation} where
$v_i^k$ is the $i$ component of $v_F$ on the sheet $k$ ($k\in \sigma,\pi$), and the integral is over each pair of Fermi
surface sheets.
 In MgB$_2$ the energy gaps are expected to deviate somewhat from the usual BCS $T$ dependence
\cite{GolubovBDKJ02,Kogan02}. A simple correction \cite{CarringtonM03} to the two gap model is to use the temperature
dependent gap functions as calculated using an anisotropic strong coupling model \cite{BrinkmanGRDKKJA02} (see inset
Fig.~\ref{figrho}).

Fits of data to the two-gap model, calculated using $\Delta_{k}(T)$ from Ref. \onlinecite{BrinkmanGRDKKJA02} are shown
in Fig.\ \ref{figrho}. As discussed previously \cite{ManzanoCHLYT02,CarringtonM03}, fits to $\rho_a(T)$ give
$\Delta_\pi = 29\pm 2$~K and $\Delta_\sigma = 75 \pm 5 $~K, which are in good agreement with values obtained from other
measurements. The relative weight on each band $x_a = 0.53\pm 0.04$, is consistent with Eq.\ (\ref{xeq}) using the
calculated Fermi surface parameters \cite{Kogan02}.

The strong anisotropy of the $\sigma$ sheets should result in the $c$-axis response being dominated by the more
isotropic $\pi$ sheets \cite{Kogan02}. It can be seen in Fig.\ \ref{figrho} that $\rho_c(T)$ is quite different from
$\rho_a(T)$ in a manner consistent with the smaller contribution from the $\sigma$ band, and appears to be mostly
determined by the $\pi$ band where the gap is smaller. A fit of the data to the two gap model is shown in the figure.
Here we have fixed $\Delta_\sigma$ and $\Delta_\pi$ to the values found from the fit to $\rho_a(T)$, and have set
$x_c=0.91$ as this produces the correct value for the measured anisotropy, $\gamma_\lambda$ at $T_c$ (see below). The
fit does not change appreciable if we allow $x_c$ to vary by $\pm$ 10\%, and so is compatible which that expected from
the band structure ($x_c$=0.99) (i.e., the contribution from the $\sigma$ band is not discernable).

The fit is slightly worse than for the in-plane data, and does not improve markedly if we allow the
values of the adjustable parameters ($x_c$, $\Delta_\sigma$ and $\Delta_\pi$) to vary within
acceptable limits. We also find the although the exact $T$ dependence of $\Delta$ used in the model
has little effect on the calculations of $\rho_a(T)$ \cite{CarringtonM03}, $\rho_c(T)$ and
$\gamma_\lambda$ are somewhat more sensitive. This suggests that insufficient accuracy in the
assumed form of $\Delta_\pi(T)$ are responsible for these discrepancies.

\begin{figure}
\center
\includegraphics*[width=6cm]{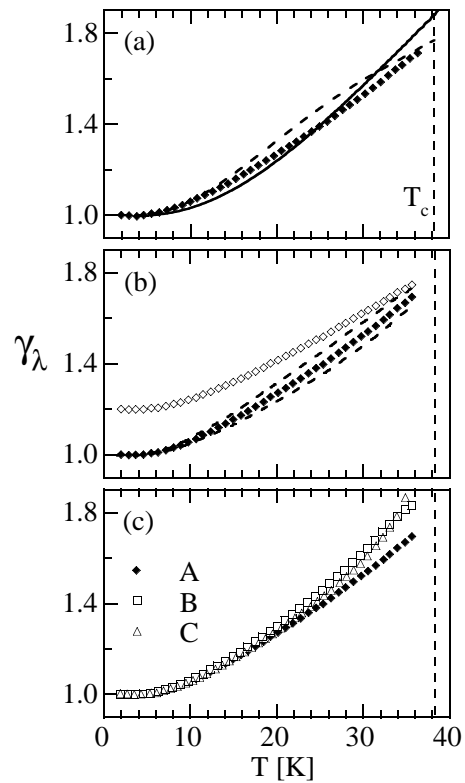}
\caption{(a) Penetration depth anisotropy $\gamma_\lambda$ versus temperature for sample A (symbols).  The solid/dashed
lines show the behavior calculated by Golubov \textit{et al.} \cite{GolubovBDKJ02} and the phenomenological two gap
model respectively.(b) $\gamma_\lambda(T)$ for sample A calculated with either $\gamma_\lambda(0)=1.0$ (solid symbols)
or $\gamma_\lambda(0)=1.2$ (open symbols). The dashed lines show the effect of changing $\lambda_{a}(0)$ from 1000~\AA~
to 800~\AA~(upper curve) or 1200~\AA~(lower curve)[with $\gamma_\lambda(0)=1.0$]. (c) $\gamma_\lambda(T)$ for three
different samples (A,B,C).}
 \label{figglam}
\end{figure}

From our measurements of $\Delta\lambda_{a}$ and $\Delta\lambda_c$ we can calculate the anisotropy in the penetration
depth as a function of temperature. This is shown in Fig.\ \ref{figglam}(a), using the same values of $\lambda_a(0)$
and $\lambda_c(0)$ as used above. Below $T\simeq7$~K, $\gamma_\lambda$ like $\Delta\lambda_a(T)$ and
$\Delta\lambda_c(T)$ is quite temperature independent, but above this it rises monotonically reaching a value of
1.7$\pm0.3$ at T$_c$. Changing the value of $\lambda_{a}(0)$ from 800 to 1200~\AA~or changing $\gamma_\lambda(0)$ in
the range 1.0 to 1.2 does not modify the $T$ dependence of $\gamma_\lambda$ significantly and also has very little
effect on the limiting value of $\gamma_\lambda$ at $T_c$ [see Fig.~\ref{figglam}(b)]. The main uncertainty in
$\gamma_\lambda(T)$ comes from the calculation of calibration factors used to extract $\Delta\lambda_a(T)$ and
$\Delta\lambda_c(T)$ from the measured frequency shifts. To obtain independent estimates of the errors we have made
measurements on two other crystals with different aspect ratios ($w/t$ ranges from 8 to 2.5). These are shown in Fig.
\ref{figglam}(c). For crystal B, $\gamma_\lambda$ is $\sim$10\% larger at $T_c$ but has a very similar $T$ dependence
to crystal A. Crystal C is again similar but shows more upward curvature near $T_c$.  We note that close to $T_c$ the
results are particularly sensitive to any macroscopic inhomogeneity in the crystal (giving rise to a spread in $T_c$
values). Data for $T/T_c>0.95$ are particularly unreliable for this reason and are omitted from the figure.

In Fig.\ \ref{figglam}(a) we show that our data are in good agreement with the predication of the (parameter free)
strong coupling calculations of Golubov \textit{et al.} \cite{GolubovBDKJ02} (in the clean limit). However, it should
be mentioned that Ref.\ \cite{GolubovBDKJ02} predicts $\lambda(0)$ values which are $\sim$ 2 times smaller than the
experimental ones for samples of the purity we have here.  We also show in this figure the calculated anisotropy of the
two gap model based on the fits to the data in Fig.\ \ref{figrho}. As mentioned above, we have adjusted $x_c$ to give
the correct value of anisotropy at $T_c$.  As for the fits to $\rho_c(T)$, the agreement is reasonable.

\begin{figure}
\center
\includegraphics*[width=8cm]{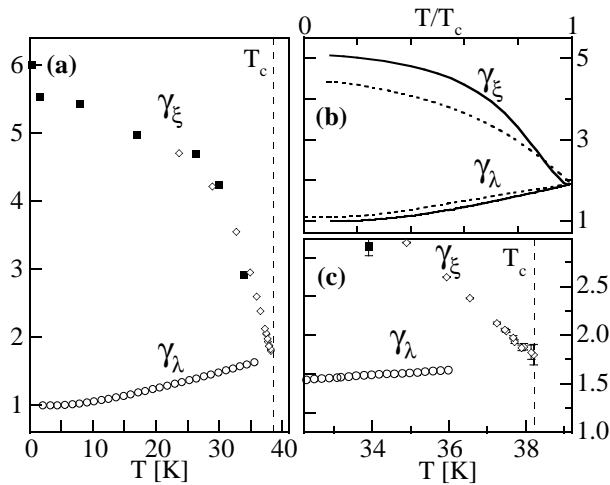}
\caption{(a) Temperature dependence of $\gamma_\xi$ and $\gamma_\lambda$.  For $\gamma_\xi$ the open symbols are
specific heat measurements and the closed symbols torque measurements. The $\gamma_\lambda$ data are the same as in
Fig.\ \ref{figglam}a. (b) Theoretical predictions for $\gamma_\xi(T)$ and $\gamma_\lambda(T)$; solid lines --
Golubov/Rydh \cite{GolubovBDKJ02,RydhWKKCBLKMKKKLL04},  dashed lines -- Miranovic/Kogan \cite{MiranovicMK03}. (c)
Detail of the experimental data in (a) close to $T_c$.} \label{figghc2}
\end{figure}

For completeness we have also measured the anisotropy of $H_{c2}$. This was done using both torque and specific heat
measurements. The heat capacity measurements were performed on the same sample (sample A) as that used for the
$\lambda(T)$ measurements. For the torque measurements a different, smaller sample was used.  The heat capacity $c_p$
was measured as a function of temperature by an a.c.\ technique \cite{CarringtonMBCBMH97} in fixed fields up to 7~T
applied parallel or perpendicular to $c$. The jump in $c_p$ at $T_c(H)$, was used to deduce $H_{c2}(T)$ in the two
directions. In order to calculate $\gamma_\xi(T)$, we need to know $H_{c2}$ in both field directions at a single
temperature. As $H_{c2}^{\|c}$ is almost linear with $T$ for $H$ approximately parallel to $c$ we interpolated this
data set to give $H_{c2}^{\|c}(T)$ at the same temperature as we have data for $H\bot c$.  Torque was measured as a
function of $H$ and angle at various temperatures between 0.3~K and 35~K using a piezoresistive cantilever technique
\cite{RosselBZHWK96}. The angular dependence of $H_{c2}$ was then fitted to the anisotropic Ginzburg-Landau form to
give $\gamma_\xi$. Extracting $H_{c2}$ from the torque measurements \cite{AngstPWJKKRK02,Fletcher} is more difficult
than for heat capacity as pinning effects, especially at low temperature, complicate the behavior close to $H_{c2}$. We
find however, that in the region of overlap there is good agreement between the two techniques. Our results for
$\gamma_\xi(T)$ shown in Fig.\ \ref{figghc2} are very similar to those reported previously \cite{RydhWKKCBLKMKKKLL04}.
In Fig.\ \ref{figghc2} we show our data for both $\gamma_\xi(T)$ and $\gamma_\lambda(T)$.  Close to $T_c$ it can be
seen that both anisotropies converge to a single value.  The $c_p$ measurements are particularly well suited to
determining the behavior of $\gamma_\xi$ close to $T_c$ and we find that $\gamma_\xi(T_c)=1.75\pm0.05$.

In Fig.\ \ref{figghc2}(b) we show two different theoretical predictions for both $\gamma_\xi(T)$ and
$\gamma_\lambda(T)$.  The solid lines show the results of the strong coupling calculations [$\gamma_\lambda(T)$ is as
in Fig.\ \ref{figglam} and $\gamma_\xi(T)$ is the theory of Golubov and Koshelev \cite{GolubovK03} using the fitting
parameters determined by Rydh \textit{et al.} Ref.\ \cite{RydhWKKCBLKMKKKLL04}. The dashed lines show the calculations
of Miranovic and Kogan \cite{MiranovicMK03} (MK) reevaluated for a gap ratio $\Delta_\sigma/\Delta_\pi = 2.7$.  Both
these calculations are in good overall agreement with our data. The MK calculation underestimates the low $T$ value of
$\gamma_\xi$ because of the assumed Fermi surface shape (elliptical rather than cylindrical) but correctly predicts
that $\gamma_\xi \simeq \gamma_\lambda \simeq 1.9$ at $T_c$. Although the calculations of Golubov \textit{et al.} for
$\gamma_\lambda(T)$ are parameter free, those for $\gamma_\xi$ are not and $\gamma_\xi$ and $\gamma_\lambda$ are only
equal at $T_c$ for certain parameter values.

In conclusion, we have measured the temperature dependent anisotropy of both the penetration depth and $H_{c2}$ in
MgB$_2$. As predicted by theory, they have very different temperature dependencies but tend to a common value at $T_c$.

We thank A.\ Golubov, A.\ Koshelev,  P.\ Miranovic and V.\ Kogan for giving us details of their calculations.


\end{document}